\documentclass[12pt]{article}
\usepackage{amssymb,amsfonts,times,graphics,psfrag}
\usepackage[mathscr]{eucal}
\renewcommand{\baselinestretch}{1.2}
\makeatletter

\@addtoreset{equation}{section}
\renewcommand\tableofcontents{
 \begin{center}\bf\Large\contentsname\end{center}
 \@starttoc{toc}
}
\makeatother

\newcommand{\nn}{\nonumber}
\newcommand{\ts}{\textstyle}
\newcommand{\ket}[1]{\vert  #1\rangle}
\newcommand{\vev}[1]{\left\langle{#1}\right\rangle}

\newcommand{\KK}{{\mathbb K}}
\newcommand{\cB}{{\cal B}}
\newcommand{\cO}{{\cal O}}
\newcommand{\cT}{{\cal T}}
\newcommand{\cX}{{\cal X}}
\newcommand{\cY}{{\cal Y}}
\newcommand{\bG}{{\bf G}}
\newcommand{\bS}{{\bf S}}

\newcommand{\cgh}{{\rm c}}
\newcommand{\bgh}{{\rm b}}

\begin{document}

\pagestyle{empty}
\baselineskip 5mm
\hfill
\hbox to 3cm
{\parbox[t]{5cm}{KIAS-P08033} \hss}\\

\baselineskip0.8cm\vskip2cm
\vskip10mm
\begin{center}

 {\Large\bf Minimal Open Strings}

\end{center}
\vskip10mm
\baselineskip0.6cm
\begin{center}

    Kazuo Hosomichi
 \\ \vskip2mm
{\it Korea Institute for Advanced Study\\ Seoul 130-012, Korea} \vskip3mm

\end{center}
\vskip8mm\baselineskip=3.5ex
\begin{center}{\bf Abstract}\end{center}\par\smallskip

We study FZZT-branes and open string amplitudes in $(p,q)$ minimal
string theory.
We focus on the simplest boundary changing operators in two-matrix
models, and identify the corresponding operators in worldsheet
theory through the comparison of amplitudes.
Along the way, we find a novel linear relation among FZZT
boundary states in minimal string theory.
We also show that the boundary ground ring is realized on physical
open string operators in a very simple manner, and discuss its use
for perturbative computation of higher open string amplitudes.

\vspace*{\fill}
\noindent April 2008
\newpage
\setcounter{page}{0}

\pagestyle{plain}
\pagenumbering{arabic}

\section{Introduction}\label{sec:Intro}

Non-critical strings propagating in low-dimensional space-time
are interesting toy models of strings
\cite{Douglas:1989ve}--\cite{Fukuma:1996hj}.
There are very few dynamical degrees of freedom in such models, and
the dynamics is heavily constrained by a large symmetry or integrability.
Also, it has long been known that these models have dual non-perturbative
descriptions in terms of large $N$ matrix models.

Recent developments in string theories have lead us to realize
that D-branes are present also in non-critical string theories.
Since a big breakthrough was made in the study of Liouville
theory on worldsheets with boundary
\cite{Fateev:2000ik}--\cite{Ponsot:2001ng},
many earlier results from matrix models have been revisited and
combined with the modern ideas
\cite{McGreevy:2003kb}--\cite{Fukuma:2006ny}.
This has brought us with a much deeper insight into the models.
Now that we have a rather precise understanding of D-branes, it is
natural to go further to study the dynamics of open strings.
In particular it will be interesting to study how much of the
open string dynamics is governed by symmetry.

In this note we wish to study some simple open string amplitudes
in $(p,q)$ minimal string theory.
We will study them from two different frameworks; using the two-matrix
model in Section \ref{sec:mm} and the worldsheet $(p,q)$ minimal model
coupled to Liouville theory in Section \ref{sec:wst}.
Along the way, we find a curious linear relation among FZZT boundary
states in the worldsheet theory of $(p,q)$ minimal string.
In Section \ref{sec:bgr} we compute the action of boundary ground
ring on physical open string operators and discuss its possible
application to higher point amplitudes.

\section{Matrix Model}\label{sec:mm}

It is known that the minimal string theories can be formulated
as large $N$ matrix integrals.
Throughout this paper we will use the two-matrix model.
We begin with reviewing the definition and some fundamental results
of this model.
See \cite{Ginsparg:1993is, Di Francesco:1993nw} for more detail.

Two-matrix model \cite{Brezin:1989db}--\cite{Daul:1993bg} is an
integral over two $N\times N$ Hermitian matrices $X,Y$:
\begin{equation}
 \int dXdY \exp\left[-\frac Ng{\rm Tr}\left(V(X)+U(Y)-XY\right)\right].
\end{equation}
We assume that $V(X)$ and $U(Y)$ are polynomials of degree $q$ and $p$,
the simplest choice for realizing $(p,q)$ critical behavior.
Standard Feynman graph expansion allows us to express the partition
function as a sum over fishnet diagrams of arbitrary area (number of
vertices) and topology.
Each diagram is regarded as a two-dimensional Riemann surface painted
by two colors `X' and `Y'.
The contribution from genus $h$ diagram is proportional to $N^{2-2h}$,
so $1/N$ plays the role of bare string coupling.

After using Harish-Chandra-Itzykson-Zuber formula to reduce the
integral to that over the eigenvalues, one is lead to consider
the set of polynomials $\{\psi_n(x),\tilde\psi_n(y)\}$ satisfying
\begin{equation}
 \int dxdy e^{-\frac Ng\left[V(x)+U(y)-xy\right]}\psi_n(x)\tilde\psi_m(y)
 ~=~ \delta_{nm}.
\end{equation}
The indices $n,m$ represent the degree of the polynomials.
The two matrices then turn into operators $\hat X$ and $\hat Y$
acting on the set of polynomials as multiplications by $x$ or $y$.
The exact partition function of two-matrix model can be expressed
in terms of the matrix elements of $\hat X$ and $\hat Y$.

\paragraph{Spectral curve.}
A fundamental observable is the resolvent,
\begin{equation}
 R_X(x)\equiv{\rm Tr}\frac{1}{X-x},~~~~~
 R_Y(y)\equiv{\rm Tr}\frac{1}{Y-y}.
\end{equation}
They carry the important information on the eigenvalue distributions.
Classically at $g=0$ each pair of eigenvalues $(x_i,y_i)$ sits on
one of the classical saddle points satisfying $y=V'(x), x=U'(y)$.
At nonzero $g$ the eigenvalues spread due to repulsive Coulomb force
arising from integrating out the off-diagonal matrix elements.

In the planar approximation, the two equations
\begin{equation}
 y = V'(x)+\frac gNR_X(x),~~~~~
 x = U'(y)+\frac gNR_Y(y)
\end{equation}
are known to give the same equation on $(x,y)$ defining the
{\it spectral curve}.
When regarded as a complex curve, its branch structure reflects
how the eigenvalues of $X,Y$ are distributed near each saddle point.
It is natural to find the true minimum of the classical action
and perform perturbative expansion around that ground state.
Such classical ground state should correspond to the spectral curve
which is a complex curve of genus zero.

\paragraph{Continuum limit.}

The idea to get continuous worldsheet is to send $N\to\infty$ and
$g\to g_c$ in a suitably correlated manner.
Going back to the system of orthonormal polynomials, we find
the index $n$ can be replaced by a continuous variable $z=gn/N$
at large $N$.
We parametrize the region $z\sim g_c$ by a new variable
$t\equiv\varepsilon^{-2}(g_c-z)$, put $N=\varepsilon^{\gamma-2}$
and take $\varepsilon\to0$.
For judiciously choson potentials, we find the operators
$\hat X$ and $\hat Y$, after suitable rescaling, become a pair of
differenial operators
\[
\begin{array}{rcl}
 \hat X &\sim& d^p+u(t)d^{p-2}+\cdots,\\
 \hat Y &\sim& d^q+v(t)d^{q-2}+\cdots,
\end{array}
~~~~\left(\ts d\equiv \frac{d}{dt},~\gamma=-\frac{2}{p+q-1}\right)
\]
satisfying the canonical commutation relation $[\hat X,\hat Y]=1$
or {\it string equation}.
It is known that these operators are conveniently expressed in terms
of powers $L^j$ of a pseudo-differential operator $L=d+{\cal O}(d^{-1})$,
whose positive parts $L^j_+$ generate mutually
commuting flows by $\frac{\partial}{\partial t_j}L = [L^j_+,L]$.
For $Y=L^q$, the solution to the string equation is
\begin{equation}
 X ~=~ -\sum_{j=1}^p    (1+{\ts\frac jq})t_{j+q}L^j_+
   ~=~ -\sum_{j=1}^{p+q}\frac jqt_{j}L^{j-q}+{\cal O}(d^{-1-q}).
\end{equation}
The string equation allows us to determine all the coefficient
functions ($u,v,\cdots$) and therefore the partition function
as functions of couplings $(t_1,t_2,\cdots)$.
For $p>q$, the conformal $(p,q)$ minimal string is obtained by turning on
only $t_{p+q}$ and $t_{p-q}$.
After fixing the former, the latter plays the role of the cosmological
constant.

The resolvents of two-matrix models for $(p,q)$ minimal string
were computed in \cite{Kostov:1991hn}.
The spectral curve is given by $y=R_{\hat X}(x)$ and $x=R_{\hat Y}(y)$
and has a simple parametric expression
\cite{Daul:1993bg}
\begin{equation}
 x ~=~2u^{\frac p2}\cos(\pi\theta/q),~~~~
 y ~=~2u^{\frac q2}\cos(\pi\theta/p),~~~~
 pu^q ~=~ (p-q)t_{p-q}.
\end{equation}
Here $\theta\sim\theta+2pq$ is the uniformizing parameter.
Hereafter we set $u=1$ for convenience.
Using Chebyshev polynomials $T_n(\cos\theta)=\cos n\theta$, the
spectral curve can be written in an algebraic form
\begin{equation}
 E(x,y)~\equiv~ T_q(x/2)-T_p(y/2) ~=~0.
\end{equation}

\subsection{Some disk amplitudes}\label{sec:mmamp}

The resolvent $R_{\hat X}(x)$ is related via Laplace transform
to the operator ${\rm Tr}e^{l\hat X}$ that creates a macroscopic
loop of length $l$ \cite{Moore:1991ir,Moore:1991ag}.
We define the FZZT boundary condtion in minimal string theory by
weighting each macroscopic loop of length $l$ by a factor
$e^{-lx}$, where $x$ is called the boundary cosmological constant.
To the leading order in large $N$, the correlator
\begin{equation}
 -\vev{{\rm Tr}\log(\hat X-x)}~=~ \int\frac{dl}{l}\vev{{\rm Tr}e^{l(\hat X-x)}}
\end{equation}
gives the disk partition function.
The resolvent is its first $x$-derivative so that it has one insertion of
boundary cosmological operator $\cB$ along the loop.
Using the uniformization coordinate $\theta$,
\begin{equation}
 \vev{{}^\theta[\cB]^\theta}
 ~=~ \vev{{\rm Tr}\frac{1}{{\hat X}-x(\theta)}}
 ~=~ y(\theta).
\end{equation}

When there are more than one insertions of $\cB$, one may assign
different boundary cosmological constants to each boundary segment.
Such amplitudes are the simplest amplitudes of open strings stretching
between different FZZT-branes.
We can compute them by the iterative use of the simple formula
\[
 \frac{1}{(\hat X-x_1)(\hat X-x_2)}~=~
 \frac{1}{x_1-x_2}\left(\frac{1}{\hat X-x_1}-\frac{1}{\hat X-x_2}\right).
\]
Explicitly, one finds
\begin{eqnarray}
 \vev{{}^{\theta_1}[\cB]^{\theta_2}[\cB]^{\theta_1}}
 &=& \vev{{\rm Tr}\frac{1}{({\hat X}-x_1)({\hat X}-x_2)}}
 ~=~ \frac{y_1-y_2}{x_1-x_2},
 \label{mm2pt} \\
 \vev{{}^{\theta_1}[\cB]^{\theta_2}[\cB]^{\theta_3}[\cB]^{\theta_1}}
 &=& \vev{{\rm Tr}\frac{1}{({\hat X}-x_1)({\hat X}-x_2)({\hat X}-x_3)}}
 \nn\\
 &=& \frac{x_1y_2+x_2y_3+x_3y_1-y_1x_2-y_2x_3-y_3x_1}
          {(x_1-x_2)(x_2-x_3)(x_3-x_1)},
\label{mm3pt}
\end{eqnarray}
where $x_i=x(\theta_i),\; y_i=y(\theta_i)$.
General $n$-point amplitude becomes
\begin{equation}
 \vev{{}^{\theta_1}[\cB]^{\theta_2}\cdots{}^{\theta_n}[\cB]^{\theta_1}}
 ~=~
 \frac{(-)^{\frac12n(n-1)}}{\Delta(x_i)}
 {\rm det}\left(
 \begin{array}{ccccc}
 1      & x_1   & \cdots & x_1^{n-2} & y_1 \\
 \vdots &\vdots &        & \vdots    & \vdots \\
 1      & x_n   & \cdots & x_n^{n-2} & y_n \\
 \end{array} \right).
\end{equation}

A more non-trivial boundary operator is the one which changes
the color of the boundary, which we call $\cT$ in the following.
The amplitudes of such operators are given by ``mixed-trace''
correlators, and they have been extensively studied in a recent
work by Eynard, et. al. using the loop equations
\cite{Eynard:2003kf}--\cite{Eynard:2007gw}.
The simplest example is the two-point correlator, which in the planar
limit is given by \cite{Eynard:2003kf}
\begin{equation}
 \vev{{\rm Tr}\frac{1}{{\hat X}-x}\frac{1}{{\hat Y}-y}}
 ~=~ \frac{E(x,y)}{(x-R_{\hat X}(y))(y-R_{\hat Y}(x))}.
\end{equation}
As a function of $\theta,\theta'$ it becomes, up to normalization,
\begin{eqnarray}
 \vev{{}^\theta[{\cal T}]^{\theta'}[{\cal T}]^{\theta}}
 &=&
  \frac{2\cos\pi\theta-2\cos\pi\theta'}
       {\{x(\theta)-x(\theta')\}\{y(\theta)-y(\theta')\}}.
\label{mmTT}
\end{eqnarray}
Note that the enumerator can be factorized,
\begin{equation}
 2\cos\pi\theta-2\cos\pi\theta'
 ~=~ \prod_{j=0}^{q-1}\{x(\theta)-x(\theta'+2pj)\}
 ~=~ \prod_{j=0}^{p-1}\{y(\theta)-y(\theta'+2qj)\} .
\end{equation}
Disk amplitudes containing more ${\cal T}$'s can be computed using the
recursion relation of \cite{Eynard:2005iq}.

\section{Worldsheet Theory}\label{sec:wst}

The worldsheet theory of $(p,q)$ minimal string is the product of
a Liouville theory with $b=\sqrt{p/q}$ and a $(p,q)$ minimal model.
In this section we generalize this and study the product of two
Liouville theories with the couplings $b$ and $ib$
\cite{Di Francesco:1991ud,Kostov:2005av}.
We start with reviewing the Liouville theory in the presence of
boundary.

\subsection{Liouville theory with boundary}\label{sec:ltwb}

Liouville theory with coupling $b$ is a theory of a scalar field $\phi$
with a potential $\mu e^{2b\phi}$.
It is a CFT with central charge
\begin{equation}
 c ~=~ 1 + 6Q^2~~~~~(Q=b+b^{-1}).
\end{equation}
Boundary conditions of Liouville theory are classified by
\cite{Fateev:2000ik, Zamolodchikov:2001ah}.
Some of them, called FZZT boundary states, are described by the boundary
interaction $\mu_B \cB$, where the cosmological operator
$\cB\equiv\oint e^{b\phi}$ measures the length of the boundary.
We parametrize the boundary states by $s$, in terms of which
$\mu_B$ is given by
\[
 \mu_B ~=~ x(s) ~\equiv~ \sqrt{\mu\pi\gamma(b^2)}
           \times\frac{\Gamma(1-b^2)}{\pi}\cos(\pi bs).
\]
In the following we set $\mu\pi\gamma(b^2)=1$ by a suitable
constant shift of the Liouville field.
The dual boundary cosmological constant $y(s)$ is related to
$x(s)$ by $b\leftrightarrow 1/b$ flip.

Boundary operator $B_k=e^{\frac{(Q+k)\phi}2}$ has weight
$\frac{Q^2-k^2}{4}$ and satisfies reflection relation
\begin{equation}
  {}^s[B_k]^t ~=~ {}^s[B_{-k}]^t\times d(k,s,t).
\label{Lref}
\end{equation}
The coefficient $d(k,s,t)$ is given by
\begin{equation}
 d(k,s,t) ~=~ \ts
 \bG(-k)\bG(k)^{-1}b^{kb-\frac kb}
 \bS(\frac{Q-k+s+t}2)
 \bS(\frac{Q-k+s-t}2)
 \bS(\frac{Q-k-s+t}2)
 \bS(\frac{Q-k-s-t}2) .
\end{equation}
Here the functions $\bG(x)$ and $\bS(x)=\bG(Q-x)/\bG(x)$ are the special
functions introduced in \cite{Fateev:2000ik}.
They are characterized by the shift equations
\begin{equation}
\begin{array}{rcl}
 \bS(x+b) &=& 2\sin(\pi bx)\bS(x),\\
 \bS(x+\frac1b) &=& 2\sin(\pi x/b)\bS(x),
\end{array}
~~~~
\begin{array}{rcl}
 \bG(x+b) &=& (2\pi)^{-\frac12}b^{\frac12-bx}\Gamma(bx)\bG(x), \\
 \bG(x+\frac1b) &=& (2\pi)^{-\frac12}b^{\frac xb-\frac12}\Gamma(x/b)\bG(x).
\end{array}
\end{equation}
As a special case, we have
\begin{equation}
 d(b-{\ts\frac1b},s,t) ~=~
  \frac{y(s)-y(t)}{x(s)-x(t)}.
\end{equation}

\paragraph{Degenerate operators.}

The boundary operators $B_k$ with special $k$ correspond to
degenerate representations.
They are used to construct the boundary ground ring elements
in minimal string theory.
The basic ones are $X\equiv e^{-\frac{b\phi}{2}}$ and
$Y\equiv e^{-\frac{\phi}{2b}}$.
$X$ or $Y$ are known to connect two boundary states whose $s$ labels
differ by $\pm b$ or $\pm b^{-1}$, respectively.
Their OPEs with general boundary operators read \cite{Fateev:2000ik},
\begin{eqnarray}
 {}^{s'}[X(z)]^s[B_k(w)]^t &=&
 \sum_\pm X_\mp\;|z-w|^{\frac b2(Q\pm k)}\cdot{}^{s'}[B_{k\mp b}(w)]^t,
 \nn\\
 {}^{s'}[Y(z)]^s[B_k(w)]^t &=&
 \sum_\pm Y_\mp\;|z-w|^{\frac 1{2b}(Q\pm k)}\cdot{}^{s'}[B_{k\mp\frac1b}(w)]^t.
\label{XBope}
\end{eqnarray}
The coefficients are given by $X_-=Y_-=1$ and
\begin{eqnarray}
 X_+ &=& \ts
 \frac{2b^2}{\pi}\Gamma(-bk-b^2)\Gamma(bk)
 \sin\pi(\frac{b(Q+k\pm s+t)}{2})
 \sin\pi(\frac{b(Q+k\pm s-t)}{2})
~~~~(s'=s\pm b), \nn\\
 Y_+ &=& \ts
 \frac{2}{\pi b^2}
 \Gamma(-\frac kb-\frac{1}{b^2})\Gamma(\frac kb)
 \sin\pi(\frac{Q+k\pm s+t}{2b})
 \sin\pi(\frac{Q+k\pm s-t}{2b})
~~~~~~~~~~~~~(s'=s\pm \frac1b).
\label{XBope2}
\end{eqnarray}

\paragraph{The second Liouville theory.}

As the matter theory, we consider the second Liouville theory
with coupling $ib$ and the central charge
\[
 c~=~1+6\tilde Q^2 ~~~~(\tilde Q=ib-ib^{-1}).
\]
The product of Liouville theories with couplings $b$ and $ib$
has critical central charge.
We put a tilde to every quantity in the second Liouville theory:
for example, the boundary operators $\tilde B_{ik}$ have weight
$\frac{\tilde Q^2+k^2}4$.
The basic degenerate operators are denoted by $\tilde X$ and $\tilde Y$,
and when multiplied on $\tilde B_{ik}$ they shift the momentum $k$ by
$\pm b$ or $\pm b^{-1}$.

For $b=\sqrt{p/q}$ the second Liouville theory can be reduced to
the $(p,q)$ minimal model with finitely many primary fields forming
a closed algebra under fusion.
Also, in minimal models there are finitely many boundary states (Cardy states)
corresponding to special values of the parameter $\tilde s$,
\begin{equation}
 \tilde s~\in~ \KK\equiv\{lb-kb^{-1}~|~1\le k\le p-1~,~1\le l\le q-1\}.
\label{kac}
\end{equation}
Although their property is significantly different from that of
FZZT boundary states in Liouville theory with generic $b$, the OPE formula
(\ref{XBope}), (\ref{XBope2}) should apply to them as well.
This is because the OPE coefficients appearing there are essentially
the fusion matrix elements, and they depend on the boundary conditions
only through their $s$-parameters.

\subsection{FZZT-branes}\label{sec:fzzt}

The FZZT-brane $\ket{s;k,l}$ in minimal string theory is defined as
the direct product of a FZZT boundary state in Liouville theory and
the $(k,l)$ Cardy state in minimal model.
Its Liouville part is characterized by the boundary cosmological
constant and its dual,
\begin{equation}
 x(s)=2\cos(\pi sb),~~~~
 y(s)=2\cos(\pi s/b),
\end{equation}
where some unimportant factors has been dropped.
Comparison of this with the result from matrix model shows that the
spectral curve has an interpretation as the moduli space of FZZT-branes
\cite{Seiberg:2003nm}.
The uniformization parameters in the two frameworks are related
by $\theta=s\sqrt{pq}$.

\begin{figure}[htb]
{
\centerline{
\scalebox{0.8}{
\psfrag{arccosx}{$\cos^{-1}(x/2)$}
\psfrag{arccosy}{$\cos^{-1}(y/2)$}
\includegraphics{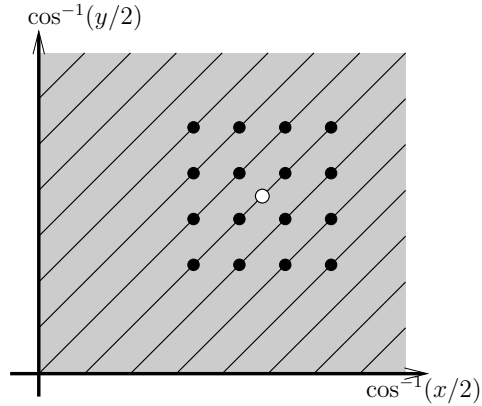}}
  } }
  \caption{\small The oblique lines form the spectral curve for the
  two-matrix model realizing $(p,q)=(8,7)$ minimal string.
  The curve covers the $x$-plane 7 times and $y$-plane 8 times.
  The white dot is an FZZT-brane $\ket{\theta;3,3}$ which decomposes
  into nine elementary FZZT-branes described by black dots.}
\label{fig:curve}
\end{figure}

Apparently, the worldsheet theory has more D-branes than the two-matrix
model, since the branes in the latter do not have labels $(k,l)$.
A proposal to resolve this mismatch has been made in
\cite{Seiberg:2003nm}: it has been observed there that the FZZT-brane
$\ket{\theta;k,l}$ in minimal string theory with $(k,l)\ne(1,1)$
is equivalent to the sum of $(k\times l)$ {\it elementary branes}
$\ket{\theta'}\equiv\ket{\theta';1,1}$,
\begin{equation}
 \ket{\theta;k,l} ~\simeq~
 \sum_{i,j} \ket{\theta+qj+pi},~~~~
 \left\{\begin{array}{rcl}
 j&\in&\{1-k,3-k,\cdots,k-1\},\\
 i&\in&\{1-l,3-l,\cdots,l-1\}.
\end{array}
 \right.
\label{bdcmp}
\end{equation}
This equivalence has been checked in \cite{Seiberg:2003nm} in the sense
of BRST cohomology, and derived in \cite{Basu:2005sda} using the
boundary ground ring.
The spectral curve and an example of FZZT-brane is described in Figure
\ref{fig:curve} which nicely encodes the representation theoretic aspect
of the $(p,q)$ minimal model.

\subsection{Some disk amplitudes}\label{sec:wsamp}

Here we consider some simple disk amplitudes in the generalized
minimal string theory with coupling $b$, whose worldsheet theory
is made of two Liouville theories with couplings $b$ and $ib$.
The basic physical boundary operators are boundary tachyons,
\begin{equation}
 \cB_k ~\equiv~ \cgh B_k\tilde B_{ik}.
\end{equation}
Here $\cgh$ is the reparametrization ghost field.
The boundary of the disk is labeled by a pair of parameters
$(s,\tilde s)$.
To get $(p,q)$ minimal string theory, we set $b=\sqrt{p/q}$ and restrict
$k$ and $\tilde s$ to take values in $\KK$ of (\ref{kac}).

\paragraph{Amplitudes of $\cB$.}
General three-point amplitudes are given by the product of
disk three-point functions for the two Liouville theories
\cite{Ponsot:2001ng}.
See also the recent work \cite{Furlan:2008za}.
Here we focus on the special case where the formula simplifies,
\begin{equation}
 \vev{{}^t[B_k]^s[B_{b-\frac1b}]^{s'}[B_k]^t} ~=~
 k\frac{d(k,s,t)-d(k,s',t)}{x(s)-x(s')}.
\end{equation}
From this we get the three-point amplitude
\begin{equation}
 \vev{{}^{(t,\tilde t)}[\cB_k]^{(s,\tilde s)}
      [\cB_{b-\frac1b}]^{(s',\tilde s)}[\cB_k]^{(t,\tilde t)}} ~=~
 k\frac{d(k,s,t)-d(k,s',t)}{x(s)-x(s')}\cdot
 \tilde d(ik,\tilde s,\tilde t).
\label{3ptsp}
\end{equation}
Notice that $\cB_{b-\frac1b}=\cB_{(1,1)}$ is nothing but the boundary
cosmological operator $\cB$.
Restricting to $(p,q)$ minimal string theory and setting
$k=\tilde s=\tilde t=b-\frac1b$, the three-point amplitude becomes
\begin{equation}
 \vev{{}^{(s_1)}[\cB]^{(s_2)}[\cB]^{(s_3)}[\cB]^{(s_1)}}
 ~=~
 \frac{p-q}{\sqrt{pq}}
 \frac{x_1y_2+x_2y_3+x_3y_1-y_1x_2-y_2x_3-y_3x_1}
      {(x_1-x_2)(x_2-x_3)(x_3-x_1)},
\end{equation}
where we used $x_i=x(s_i),\;y_i=y(s_i)$.
This is in agreement with the matrix model result (\ref{mm3pt}).

In the limit $s'\to s$ the right hand side of (\ref{3ptsp}) becomes
a derivative with respect to $x$.
We assume that one can integrate it when the operators inserted
are all within Seiberg's bound \cite{Seiberg:1990eb},
since it would lead to an inconsistency if we could always integrate
it \cite{Bourgine:2007ua}.
We thus find the two-point amplitude
\begin{equation}
 \vev{{}^{(t,\tilde t)}[\cB_k]^{(s,\tilde s)}
                       [\cB_k]^{(t,\tilde t)}} ~=~
 {\rm sgn}({\rm Re}k)k d(k,s,t) \tilde d(ik,\tilde s,\tilde t).
\end{equation}
Restricting to $(p,q)$ minimal string and $k=(1,1)$, we again find
the agreement with matrix model result (\ref{mm2pt}),
\begin{equation}
 \vev{{}^{(s_1)}[\cB]^{(s_2)}[\cB]^{(s_1)}} ~=~
 \frac{p-q}{\sqrt{pq}}\frac{y_1-y_2}{x_1-x_2}.
\end{equation}
We can integrate further and check that the one-point amplitude
agrees with the resolvent in two-matrix model.

\paragraph{Amplitudes of $\cT$.}
Next we consider the general two-point amplitude
in $(p,q)$ minimal string.
\begin{equation}
 \vev{{}^{(s';k',l')}[\cB_{(n,m)}]^{(s;k,l)}
                       [\cB_{(n,m)}]^{(s';k',l')}} ~\sim~
 \frac{|mp-nq|}{\sqrt{pq}}
 d(mb-nb^{-1},s,s').
\end{equation}
The amplitude is non-vanishing only when the representation
$(n,m)$ is allowed between two Cardy states $(k,l)$
and $(k',l')$ in minimal model.
More explicitly
\begin{equation}
\begin{array}{rcccl}
 |k-k'|+1 &\le& n &\le& {\rm min}(k+k'-1,2p-k-k'-1), \\
 |l-l'|+1 &\le& m &\le& {\rm min}(l+l'-1,2q-l-l'-1).
\end{array}
\label{Cacon}
\end{equation}
Using $\theta=s\sqrt{pq}$ and $\theta'=s'\sqrt{pq}$, 
the amplitude is proportional to
\[
 ~\sim~
 \frac{\prod_{j=0}^{n-1}\{y(\theta')-y(\theta+p(1-m)+q(1-n+2j))\}}
      {\prod_{j=0}^{m-1}\{x(\theta')-x(\theta+q(1-n)+p(1-m+2j))\}}.
\]
Comparing this with (\ref{mmTT}) one finds the correspondence
\begin{equation}
{}^\theta[{\cal T}]^{\theta'}~\sim~
{}^{(\theta;k,l)}[\cB_{(p-1,1)}]^{(\theta'-pq;p-k,l)}.
\label{TBcor}
\end{equation}

Thus we identified the two boundary operators $\cB$ and $\cT$
in two-matrix model with the boundary operators $\cB_{(1,1)}$ and
$\cB_{(p-1,1)}$ in the worldsheet theory.
These operators are both at the corner of Kac table.
One of their special properties is that, when fused with
any primary field, they produce only one primary.

Interestingly, by translating some four-point amplitudes from two-matrix
model into worldsheet theory, one finds that the amplitudes become
non-invariant under Liouville reflection of operators (\ref{Lref}).
The standard interpretation for this is that the insertion of
four or more operators is enough to deform the theory away from
the Liouville background.

\paragraph{New linear relation among D-branes.}

Note that (\ref{TBcor}) also suggests the equivalence between FZZT-branes
\begin{equation}
 \ket{\theta;k,l} ~=~ - \ket{\theta-pq;p-k,l}.
\label{breq}
\end{equation}
The minus sign is required for the equalities with different $(k,l)$
to be mutually consistent.
More interestingly, when these equalities are combined with (\ref{bdcmp}),
they give rise to simple linear relations among elementary FZZT-branes,
\begin{equation}
 0 ~=~ \sum_{j=1}^p\ket{\theta+2qj} ~=~ \sum_{j=1}^q\ket{\theta+2pj}~,
\label{bsum}
\end{equation}
which say that $p$ or $q$ elementary FZZT-branes can disappear
into nothing when placed in a suitable manner.
These equalities can be checked in the sense of BRST cohomology
in the same way as (\ref{bdcmp}) was checked.

\subsection{Boundary ground ring}\label{sec:bgr}

The worldsheet theory has boundary operators labeled by $(k,l)$,
but the two-matrix model does not seem to have corresponding boundary
changing operators.
In other matrix models such as height models
\cite{Pasquier:1986jc,Saleur:1988zx,Kostov:1991hn}, there seem to
be more boundary operators and we may make a more direct comparison
with the worldsheet theory \cite{inprogress}.
On the other hand, different boundary operators in worldsheet
theory are related by the action of boundary ground ring
\cite{Bershadsky:1992ub,Kostov:2003cy,Basu:2005sda} so that
we may well regard them as redundant.

There is a set of physical operators of ghost number zero in minimal
string theory which form the {\it ground ring}.
Here we consider the ring of boundary operators.
The ring elements $\cO_{m,n}$ are constructed from the
$(m,n)$ degenerate Liouville operator and the $(m,n)$ operator
in minimal model.
The ring is generated by the operators $\cX=\cO_{1,2}$
and $\cY=\cO_{2,1}$,
\begin{eqnarray}
  \cX &\equiv&
   \frac{1}{2b^2}
   (b^{+2}\,\bgh\cgh+L_{-1}-\tilde L_{-1})X\tilde X, \nn\\
  \cY &\equiv&
   \frac{b^2}{2}
   (b^{-2}\,\bgh\cgh+L_{-1}-\tilde L_{-1})Y\tilde Y.
\end{eqnarray}
Here $\bgh, \cgh$ are reparametrization ghosts.
The ring relation is realized linearly on the physical boundary
operators $\cB_k$.
Schematically one has
\begin{eqnarray}
\cX\cB_k &=& \sum_\pm \cX_\pm(k) \cB_{k\pm b}, \nn\\
\cY\cB_k &=& \sum_\pm \cY_\pm(k) \cB_{k\pm \frac1b}.
\label{XYB}
\end{eqnarray}
The coefficients $\cX_\pm, \cY_\pm$ can be computed using
the formulae (\ref{XBope}).
Similar formulae hold also for right multiplications.
Note that the coefficients depend on the boundary parameters
though we will suppress it for notational simplicity.
Note also that the boundary parameters $s$ and $\tilde s$ have to jump
by $\pm b$ or $\pm b^{-1}$ where $\cX$ or $\cY$ are inserted.

The linear action of $\cX,\cY$ on boundary tachyons satisfies
the following.
First, the left- and right-multiplications commute for all
pairs of operators,
\begin{equation}
 (\cX\cB)\cY ~=~ \cX(\cB\cY),~~~~
 (\cX\cB)\cX ~=~ \cX(\cB\cX),~~~~
 {\rm etc.}
\end{equation}
Also, the multiplications of an $\cX$ and a $\cY$
from the same side {\it anticommute},
\begin{equation}
 \cX\cY\cB   ~=~ -\cY\cX\cB.
\end{equation}
To simplify the formulae that follow, we introduce the notation
\begin{equation}
\begin{array}{rcl}
\cX_\pm &=&
 {}^{(s\pm b,\tilde s-b)}\cX^{(s,\tilde s)}, \\
\bar\cX_\pm &=&
 {}^{(s\pm b,\tilde s+b)}\cX^{(s,\tilde s)}, \\
\end{array}
~~~~
\begin{array}{rcl}
\cY_\pm &=&
 {}^{(s\mp\frac1b,\tilde s+\frac1b)}\cY^{(s,\tilde s)}, \\
\bar\cY_\pm &=&
 {}^{(s\mp\frac1b,\tilde s-\frac1b)}\cY^{(s,\tilde s)}.
\end{array}
\end{equation}
They can be shown to satisfy the algebraic relations
\begin{equation}
\begin{array}{rcl}
 \bar\cX_-\cX_+ - \bar\cX_+\cX_- &=&
 \sin(\pi bs)\sin(\pi b\tilde s-\pi b^2), \\
 \cX_+\bar\cX_- - \bar\cX_+\cX_- &=&
 \sin(\pi bs-\pi b^2)\sin(\pi b\tilde s), \\
 \bar\cY_-\cY_+ - \bar\cY_+\cY_- &=&
  \sin(\frac{\pi s}{b})\sin(\frac{\pi\tilde s}{b}+\frac{\pi}{b^2}), \\
 \cY_+\bar\cY_- - \bar\cY_+\cY_- &=&
  \sin(\frac{\pi s}{b}+\frac{\pi}{b^2})\sin(\frac{\pi\tilde s}{b}), \\
\end{array}
\label{algr}
\end{equation}
and commutation relations
\begin{equation}
\begin{array}{rcl}
   [\cX_+,\bar\cX_-] &=& -\sin\pi b^2\sin\pi b(\tilde s-s),\\
~  [\cX_-,\bar\cX_+] &=& -\sin\pi b^2\sin\pi b(\tilde s+s),\\
\end{array}
~~~~
\begin{array}{rcl}
   [\cY_+,\bar\cY_-] &=& \sin\frac{\pi}{b^2}\sin\frac\pi b(\tilde s-s),\\
~  [\cY_-,\bar\cY_+] &=& \sin\frac{\pi}{b^2}\sin\frac\pi b(\tilde s+s).
\end{array}
\end{equation}
All other commutators vanish, i.e.
\[
 [\cX_\pm,\bar\cX_\pm] = [\cX_+,\cX_-] =  [\bar\cX_+,\bar\cX_-] =
 [\cY_\pm,\bar\cY_\pm] = [\cY_+,\cY_-] =  [\bar\cY_+,\bar\cY_-] = 0.
\]

\paragraph{Linear relations among D-branes revisited.}

Thanks to the above simple algebraic relations, we may construct
general ring elements as simple products of generators without
worrying about the order of multiplication.
Let us now consider the $(p,q)$ minimal string theory and
restrict $k$ and $\tilde s$ to take values in $\KK$.
Let us introduce
\begin{eqnarray}
 {}^{(\theta')}[\cO_{k,l}]^{(\theta;k,l)} &=&
 {}^{(\theta')}[\cY_-^{k_-}\cY_+^{k_+}
                \cX_-^{l_-}\cX_+^{l_+}]^{(\theta;k,l)}, \nn\\
 {}^{(\theta;k,l)}[\bar\cO_{k,l}]^{(\theta')} &=&
 {}^{(\theta;k,l)}[\bar\cX_+^{l_+}\bar\cX_-^{l_-}
                   \bar\cY_+^{k_+}\bar\cY_-^{k_-}]^{(\theta')},
\end{eqnarray}
where $k,l,k_\pm,l_\pm,\theta$ and $\theta'$ satisfy
\begin{equation}
 \theta'=\theta+p(l_+-l_-)-q(k_+-k_-),~~~
 k = k_++k_-+1,~~~
 l = l_++l_-+1.
\label{theta'}
\end{equation}

These operators can be used to generalize the relations (\ref{bdcmp})
to the branes appearing on boundary segment.
The naive application of the formula to an FZZT-brane between two boundary
operators would lead to a conflict with Cardy's constraint.
The correct way is to put a suitable pair of boundary ground
ring elements at the ends of the segment.
By repeatedly using the first and third equalities in (\ref{algr}),
we find
\begin{equation}
]^{(\theta;k,l)}[ ~=~
\sum_{\theta'}
 \frac{
 ~]^{(\theta;k,l)}[\bar\cO_{k,l}]^{(\theta')}[\cO_{k,l}]^{(\theta;k,l)}[~}
 {F_{\theta'}(\theta;k,l)} ,
\label{bdcmp2}
\end{equation}
where the function $F_{\theta'}(\theta;k,l)$ is given by
\begin{eqnarray}
 F_{\theta'}(\theta;k,l) &=& (-1)^{l_++k_-}
 \prod_{j=-l_-}^{l_+}\sin{\ts\frac{(\theta+jp)\pi}{q}}
 \prod_{j=-k_-}^{k_+}\sin{\ts\frac{(\theta-jq)\pi}{p}}
 \nn\\&& \times
 \prod_{j=1}^{k_+}\sin{\ts\frac{jq\pi}{p}}
 \prod_{j=1}^{k_-}\sin{\ts\frac{jq\pi}{p}}
 \prod_{j=1}^{l_+}\sin{\ts\frac{jp\pi}{q}}
 \prod_{j=1}^{l_-}\sin{\ts\frac{jp\pi}{q}}.
\end{eqnarray}
The formula (\ref{bdcmp2}) can also be used to relate the three-point
amplitudes of general boundary operators to that of three boundary
cosmological operators, since
\[
 {}^{(\theta)}[\cO_{k,l}\cB_{(m,n)}\bar\cO_{k',l'}]^{(\theta')}
\]
should always be proportional to $\cB_{(1,1)}$ from Cardy's constraint.

\paragraph{Recursion relations for open string amplitudes.}

Using the operators $\cX,\cY$ one can derive recursion relations among
three-point amplitudes.
Omitting the dependence on boundary parameters,
one has schematically
\begin{eqnarray}
 0 &=&
 \langle \cB_{k_1}[Q_B,{\cal X}]\cB_{k_2}\cB_{k_3} \rangle
 \nn\\ &=&
 \langle (\cB_{k_1} \cX)\cB_{k_2} \cB_{k_3}\rangle
-\langle  \cB_{k_1}(\cX \cB_{k_2})\cB_{k_3}\rangle
 \nn\\ &=&
 \sum_\pm{\cal X}_\pm(k_1)
 \langle\cB_{k_1\pm b}\cB_{k_2}\cB_{k_3}\rangle
+\sum_\pm{\cal X}_{\pm}(k_2)
 \langle\cB_{k_1}\cB_{k_2\pm b}\cB_{k_3}\rangle.
 \label{bgr-rec}
\end{eqnarray}
Similar recursion relation can be shown to hold also for two-point amplitudes.
The idea to get these recursion relations is to rewrite the amplitudes
containing $Q_B$-exact operator into an integral over the boundary
of moduli space or a sum over factorized worldsheets.
The same arguments can be applied to obtain recursion relations
for higher amplitudes.

Concrete recursion relations have been proposed in $c=1$ string
theory by \cite{Kostov:2003cy} and in minimal string theory by
\cite{Basu:2005sda}, following the argument of \cite{Bershadsky:1992ub}
that the recursion relations boil down to the higher operator
product algebras such as
\[
 (\cB_{k_1}\cdots\cB_{k_n}\cX\cB_{k_{n+1}}\cdots\cB_{k_N})
 ~\longrightarrow~ \cB_{k'}.
\]
However, we do not see any obvious reason that the operator products vanish
for $N\ge 3$ in the worldsheet theory with nonzero cosmological coupling,
though it was assumed in many literature.

In a recent paper \cite{Bourgine:2007ua} the recursion relation
for four-point amplitudes in $c=1$ theory has been solved and shown
to reproduce the matrix model result.
It will be important to understand better the symmetry structure
of minimal string theory by making use of the boundary ground
ring relations in worldsheet theory and the loop equations
in two-matrix model.

\subsection*{Acknowledgment}

I would like to thank J.-E. Bourgine, I. Kostov and Y. Matsuo for useful
discussions, and also the organizers of RIKEN seminar series in March 2008
for hospitality during the completion of this work.

\vskip2mm

I would like to dedicate this work to the memory of
Prof. Alexei Zamolodchikov.

\newpage

\renewcommand{\baselinestretch}{1}

\end{document}